\documentclass{PoS}
\usepackage{amsmath}
\title{Invariants and CP violation in the 2HDM}

\ShortTitle{Invariants and CP violation in the 2HDM}

\author{\speaker{Odd Magne Ogreid}\\
        Western Norway University of Applied Sciences, Postboks 7030, N-5020 Bergen, Norway\\
        E-mail: \email{omo@hvl.no}}

\abstract{We discuss the importance of basis invariants in the general 2HDM and how these relates to masses and couplings. We also present a simple, yet powerful technique to translate parameters of the potential into combinations of masses and couplings of the theory and apply this to CP odd invariants.}

\FullConference{Corfu Summer Institute 2017 'School and Workshops on Elementary Particle Physics and Gravity'\\
		2-28 September 2017\\
		Corfu, Greece}

\newcommand{\nc}{\newcommand}
\nc{\beq}{\begin{equation}}  \nc{\eeq}{\end{equation}}
\nc{\bea}{\begin{eqnarray}}  \nc{\eea}{\end{eqnarray}}
\newcommand{\half}{{\textstyle\frac{1}{2}}}
\renewcommand{\Re}{\mbox{Re\thinspace}}
\renewcommand{\Im}{\mbox{Im\thinspace}}

\def\lcal{{\cal L}}
\def\mcal{{\cal M}}

\def\pcal{{\cal P}}

\def\thetaW{{\theta}_{\rm W}}
\begin{document}

\section{Introduction}
Multi-Higgs doublet models (NHDM) offer simple yet attractive extensions of the Standard Model due to the possibility to accomodate for extra CP violation as well as offering viable candidates for Dark Matter\cite{Branco:2011iw}. 

The doublets in an NHDM are not unique. One can combine the doublets to form new combinations using a $U(N)$ matrix. This is referred to as a change of basis. When writing down the Lagrangian of the 2HDM, one has a freedom to write it down in any basis one wishes. This freedom is parametrized by the arbitrary $U(N)$ matrix. The physics of the model cannot depend on this arbitrary choice of basis and implies that all observables of the model that can be measured in experiments, like for instance cross sections and decay rates, cannot depend on the choice of basis. This naturally leads to the study of basis invariant quantities, since any observable must be independent of the choice of basis.
In the Standard Model we know that the potential is invariant under a rephasing of the Higgs-doublet. This is exactly the $U(1)$ freedom present in the model with one doublet. None of the parameters of the potential change under this rephasing in the SM.

The simplest multi-Higgs extension of the Standard Model is the two-Higgs-doublet model (2HDM) in which there is a $U(2)$ freedom present.  Unlike the SM, the parameters of the potential do change when we use this freedom to change basis. Therefore, the study of basis invariants and basis invariant techniques is important in NHDMs, and many papers have been written on this subject\cite{Davidson:2005cw,Haber:2006ue,Haber:2010bw}. 

We start by discussing the general form of the 2HDM potential, vacuum expectation values and the transformation of these under changes of basis in section \ref{sec:Potential}. In section \ref{sec:Masses} we study the mass sector of the 2HDM in more detail, working out the transformation rules of both the rotation matrix and the squared mass matrix under a basis change. We use these to show that the scalar masses are invariant under a change of basis. Next, in section \ref{sec:Couplings} we study both scalar and gauge couplings and find that these are invariant under a basis change. We discuss systematic construction of basis invariant quantities in section \ref{sec:construction}. A powerful and simple method for translating from parameters to masses and couplings is presented in section \ref{sec:translation} and applied to CP odd invariants in section \ref{sec:invariants} before a short summary in section \ref{sec:summary}.

\section{The potential, vacuum expectation values and basis transformations}
\label{sec:Potential}
\subsection{Parametrizing the potential}
The potential of the two-Higgs-doublet model shall be parametrized in the standard fashion as
\begin{eqnarray}
V(\Phi_1,\Phi_2) &=& -\frac12\left\{m_{11}^2\Phi_1^\dagger\Phi_1
+ m_{22}^2\Phi_2^\dagger\Phi_2 
+ \left[m_{12}^2 \Phi_1^\dagger \Phi_2
+{\rm h. c.}\right]\right\} \nonumber \\
&&+ \frac{\lambda_1}{2}(\Phi_1^\dagger\Phi_1)^2
+ \frac{\lambda_2}{2}(\Phi_2^\dagger\Phi_2)^2
+ \lambda_3(\Phi_1^\dagger\Phi_1)(\Phi_2^\dagger\Phi_2)
+ \lambda_4(\Phi_1^\dagger\Phi_2)(\Phi_2^\dagger\Phi_1)\nonumber \\
&&+ \frac12\left[\lambda_5(\Phi_1^\dagger\Phi_2)^2 +{\rm h. c.} \right]
+\left\{\left[\lambda_6(\Phi_1^\dagger\Phi_1)+\lambda_7
(\Phi_2^\dagger\Phi_2)\right](\Phi_1^\dagger\Phi_2)
+{\rm h. c.}\right\}\nonumber\\
&\equiv& Y_{a\bar{b}}\Phi_{\bar{a}}^\dagger\Phi_b+\frac{1}{2}Z_{a\bar{b}c\bar{d}}(\Phi_{\bar{a}}^\dagger\Phi_b)(\Phi_{\bar{c}}^\dagger\Phi_d),
\end{eqnarray}
where the second form (which is also more compact) enables us to make the following identifications
\begin{eqnarray}
&&Y_{11}=-\frac{m_{11}^2}{2},\quad Y_{12}=-\frac{m_{12}^2}{2},\quad
Y_{21}=-\frac{(m_{12}^2)^*}{2},\quad Y_{22}=-\frac{m_{22}^2}{2},
\end{eqnarray}
and
\begin{eqnarray}
&&Z_{1111}=\lambda_1,\quad Z_{2222}=\lambda_2,\quad Z_{1122}=Z_{2211}=\lambda_3,\nonumber\\
&&Z_{1221}=Z_{2112}=\lambda_4,\quad Z_{1212}=\lambda_5,\quad Z_{2121}=(\lambda_5)^*,\nonumber\\
&&Z_{1112}=Z_{1211}=\lambda_6,\quad Z_{1121}=Z_{2111}=(\lambda_6)^*,\nonumber\\
&&Z_{1222}=Z_{2212}=\lambda_7,\quad Z_{2122}=Z_{2221}=(\lambda_7)^*.
\end{eqnarray}
The second form, where we express the potential parameters in terms of the tensors $Y_{a\bar{b}}$ and $Z_{a\bar{b}c\bar{d}}$ is more convenient in the study of invariants as one may write down simple and compact rules for how they transform under a change of basis, see section \ref{sec:construction}.
\subsection{Basis transformations and the parameters of the potential}
The potential is defined with respect to the doublets $\Phi_i$. There is, however, nothing unique with these (initial) doublets, and the potential may just as well be expressed in terms of the transformed doublets $\bar{\Phi}_i$, which is related to the $\Phi_i$ by $\bar{\Phi}_i=U_{ij}\Phi_j$. The matrix $U$ can be any unitary 2x2 matrix. This is what is referred to as a change of basis. Our theory can be formulated in terms of the doublets in any basis we wish. Each choice of the matrix $U$ brings us to another basis, so there are infinitely many different bases in which the theory can be formulated. Naturally, physics cannot depend on an arbitrary choice of basis, so quantities that do not depend on the matrix $U$ play an important role in the theory, and shall be referred to as {\em physical} or {\em basis invariants} of the theory.

The most general $U(2)$ matrix can be parametrized as
\begin{eqnarray}
U=
e^{i\psi}\left(
\begin{array}{cc}\cos\theta & e^{-i\xi}\sin\theta\\ -e^{i\chi}\sin\theta & e^{i(\chi-\xi)}\cos\theta
\end{array}\right).
\end{eqnarray}
Naturally, the parameters of the potential will change under a basis transformation. The transformation rules are given explicitly in \cite{Gunion:2005ja}, we repeat them here for the reader's  convenience,
\begin{eqnarray}
\bar{m}_{11}^2&=&m_{11}^2\cos^2\theta+ m_{22}^2\sin^2\theta+\Re(m_{12}^2e^{i\xi}) \sin2\theta,\\
\bar{m}_{22}^2&=&m_{11}^2\sin^2\theta+ m_{22}^2\cos^2\theta-\Re(m_{12}^2e^{i\xi}) \sin2\theta,\\
\bar{m}_{12}^2&=&\left[\half(-m_{11}^2+m_{22}^2)\sin2\theta+\Re(m_{12}^2e^{i\xi})\cos2\theta+i\Im(m_{12}^2e^{i\xi})\right]e^{-i\chi},\\
\bar{\lambda}_1&=&\lambda_1\cos^4\theta+\lambda_2\sin^4\theta+\half\lambda_{345}\sin^2 2\theta+2\sin2\theta\Re[\cos^2\theta\lambda_6e^{i\xi}+\sin^2\theta\lambda_7e^{i\xi}],\\
\bar{\lambda}_2&=&\lambda_1\sin^4\theta+\lambda_2\cos^4\theta+\half\lambda_{345}\sin^2 2\theta-2\sin2\theta\Re[\sin^2\theta\lambda_6e^{i\xi}+\cos^2\theta\lambda_7e^{i\xi}],\\
\bar{\lambda}_3&=&\frac{1}{4}\sin^2 2\theta(\lambda_1+\lambda_2-2\lambda_{345})+\lambda_3-\sin 2\theta \cos 2\theta\Re[(\lambda_6-\lambda_7)e^{i\xi}],\\
\bar{\lambda}_4&=&\frac{1}{4}\sin^2 2\theta(\lambda_1+\lambda_2-2\lambda_{345})+\lambda_4-\sin 2\theta \cos 2\theta\Re[(\lambda_6-\lambda_7)e^{i\xi}],\\
\bar{\lambda}_5&=&\left(\frac{1}{4}\sin^2 2\theta(\lambda_1+\lambda_2-2\lambda_{345})+\Re(\lambda_5 e^{2i\xi})+i\cos 2\theta \Im(\lambda_5 e^{2i\xi})\right.\nonumber\\
&&\left.\phantom{\frac{1}{4}}\hspace*{1cm}-\sin 2\theta \cos 2\theta \Re[(\lambda_6-\lambda_7)e^{i\xi}]-i\sin 2\theta \Im[(\lambda_6-\lambda_7)e^{i\xi}]\right)e^{-2i\chi},\\
\bar{\lambda}_6&=&\left(-\half\sin2\theta[\lambda_1\cos^2\theta-\lambda_2\sin^2\theta-\lambda_{345}\cos 2\theta-i \Im(\lambda_5 e^{2i\xi})]\right.\nonumber\\
&&\hspace*{1cm}+\cos\theta\cos3\theta\Re(\lambda_6e^{i\xi})+\sin\theta\sin3\theta\Re(\lambda_7e^{i\xi})\nonumber\\
&&\phantom{\half}\left.\hspace*{1cm}+i\cos^2\theta\Im(\lambda_6e^{i\xi})+i\sin^2\theta\Im(\lambda_7e^{i\xi})\right)e^{-i\chi},\\
\bar{\lambda}_7&=&\left(-\half\sin2\theta[\lambda_1\sin^2\theta-\lambda_2\cos^2\theta+\lambda_{345}\cos 2\theta+i \Im(\lambda_5 e^{2i\xi})]\right.\nonumber\\
&&\hspace*{1cm}+\sin\theta\sin3\theta\Re(\lambda_6e^{i\xi})+\cos\theta\cos3\theta\Re(\lambda_7e^{i\xi})\nonumber\\
&&\phantom{\half}\left.\hspace*{1cm}+i\sin^2\theta\Im(\lambda_6e^{i\xi})+i\cos^2\theta\Im(\lambda_7e^{i\xi})\right)e^{-i\chi}.
\end{eqnarray}
Even though the parameters of the potential change under a basis transformation, some combinations of potential parameters remain unchanged under a basis transformations. A few simple examples that can be readily verified from the above rules are
\begin{eqnarray}
\bar{m}_{11}^2+\bar{m}_{22}^2&=&m_{11}^2+m_{22}^2,\label{ind1}\\
\bar{\lambda}_1+\bar{\lambda}_2+2\bar{\lambda}_3&=&\lambda_1+\lambda_2+2\lambda_3,\label{ind2}\\
\bar{\lambda}_1+\bar{\lambda}_2+2\bar{\lambda}_4&=&\lambda_1+\lambda_2+2\lambda_4.\label{ind3}
\end{eqnarray}
These three simple basis invariants are all linear in the parameters of the potential, and they represent physical quantities of the theory due to the fact that they are basis invariant.
\subsection{Vacuum expectation values and basis transformations}
For a physically viable theory, the electroweak symmetry must be spontaneously broken so that we get non-zero vacuum expectation values (VEVs), representing the minimum of the potential. These in turn give rise to masses of the particles. The most general form of the VEVs that conserves electric charge is
\begin{equation}
\langle \Phi_1 \rangle=\frac{1}{\sqrt{2}}\left(
\begin{array}{c}0\\
v_1e^{i\xi_1}

\end{array}\right),\quad
\langle \Phi_2 \rangle =\frac{1}{\sqrt{2}}\left(
\begin{array}{c}0\\
v_2e^{i\xi_2}
\end{array}\right),
\end{equation}
where $v_1^2+v_2^2=v^2=(246\, {\rm GeV})^2$. We shall  introduce the notation $\xi_{21}\equiv\xi_2-\xi_1$ for the difference in phase between the two doublets. If we transform to a new basis, the VEVs will also change. We find that the parameters of the VEVs transform as
\begin{eqnarray}
\bar{v}_1&=&\sqrt{v_1^2\cos^2\theta+v_2^2\sin^2\theta+v_1v_2\sin2\theta\cos(\xi_{21}-\xi)},\\
\bar{v}_2&=&\sqrt{v_1^2\sin^2\theta+v_2^2\cos^2\theta-v_1v_2\sin2\theta\cos(\xi_{21}-\xi)},
\end{eqnarray}
and
\begin{eqnarray}
\cos\bar{\xi}_{21}&=&\frac{\bar{v}_1(2v_1v_2(\cos2\theta\cos(\xi_{21}-\xi)\cos\chi-\sin(\xi_{21}-\xi)\sin\chi)+(v_2^2-v_1^2)\sin2\theta\cos\chi)}
{\bar{v}_2(v_1^2+v_2^2-(v_2^2-v_1^2)\cos2\theta+2v_1v_2\cos(\xi_{21}-\xi)\sin2\theta)},\nonumber\\ \\
\sin\bar{\xi}_{21}&=&\frac{\bar{v}_1(2v_1v_2(\cos2\theta\cos(\xi_{21}-\xi)\sin\chi+\sin(\xi_{21}-\xi)\cos\chi)+(v_2^2-v_1^2)\sin2\theta\sin\chi)}
{\bar{v}_2(v_1^2+v_2^2-(v_2^2-v_1^2)\cos2\theta+2v_1v_2\cos(\xi_{21}-\xi)\sin2\theta)},\nonumber\\
\end{eqnarray}
and we immediately see that $\bar{v}_1^2+\bar{v}_2^2=v_1^2+v_2^2$. Thus, $v^2$ must represent a physical quantity of the theory that can be observed in experiments, as is of course already well known.
With the form of the VEVs given above, demanding that the potential should possess a minimum for these VEVs, one finds the stationary-point equations given in eqs. (A.1) - (A.3) of \cite{Grzadkowski:2014ada}. Enforcing that the stationary point is indeed a minimum is done by enforcing that the squared masses of the physical scalars (to be encountered in the next section) are positive so that the potential has the needed curvature for the VEVs to represent a minimum. 
\section{The physical scalars and basis invariance of the masses}
\label{sec:Masses}
We follow the same approach as in \cite{Grzadkowski:2014ada} and parametrize the doublets as 
\begin{equation}
\Phi_j=e^{i\xi_j}\left(
\begin{array}{c}\varphi_j^+\\ (v_j+\eta_j+i\chi_j)/\sqrt{2}
\end{array}\right), \quad
j=1,2.
\end{equation}
In the charged sector we extract the massless Goldstone bosons by introducing orthogonal states,
\begin{equation}
\left(
\begin{array}{c}G^\pm\\ H^\pm
\end{array}\right)
=
\left(
\begin{array}{cc}v_1/v & v_2/v\\ -v_2/v & v_1/v
\end{array}\right)
\left(
\begin{array}{c}\varphi_1^\pm\\ \varphi_2^\pm
\end{array}\right).
\end{equation}
Here, $G^\pm$ represent the massive Goldstone fields while $H^\pm$ represents the massive charged scalars. Their masses are found to be given by
\bea
M_{H^\pm}^2=\frac{v^2}{2v_1v_2\cos\xi_{21}}\Re\left(m_{12}^2-v_1^2\lambda_6-v_2^2\lambda_7-v_1v_2\left[\lambda_4\cos\xi_{21}+\lambda_5e^{i\xi_{21}}\right]\right).
\eea
Replacing each of the parameters in this expression with its basis-transformed counterpartner (i.e. putting a bar over each of the parameters) and applying the transformation rules for changes of basis listed in the previous section, we find that the expression for $M_{H^\pm}^2$ is invariant under a change of basis. The implication of this is that the charged mass is physical and can be observed in experiments, just as we expect it to be. 

Next, we turn to the neutral sector in which we introduce orthogonal states in the same way that we did in the charged sector,
\begin{equation}
\left(
\begin{array}{c}G_0\\ \eta_3
\end{array}\right)
=
\left(
\begin{array}{cc}v_1/v & v_2/v\\ -v_2/v & v_1/v
\end{array}\right)
\left(
\begin{array}{c}\chi_1\\ \chi_2
\end{array}\right).
\end{equation}
Then $G_0$ becomes the massless neutral Goldstone boson, and the remaining mass terms can be written as 
\begin{equation}
\frac{1}{2}
\left(
\begin{array}{ccc}
\eta_1 & \eta_2 & \eta_3
\end{array}
\right)
\mcal^2
\left(
\begin{array}{c}
\eta_1 \\ \eta_2 \\ \eta_3 
\end{array}
\right).
\end{equation}
The elements of the squared mass matrix $\mcal^2$ are given explicitly in eqs. (A.5) - (A.10) of \cite{Grzadkowski:2014ada}. In order to extract the physical neutral scalars, we form combinations of the $\eta_i$ by the use of an orthogonal 3x3 matrix $R$ so that the squared mass matrix becomes diagonal,
\begin{equation} \label{Eq:R-def}
\left(
\begin{array}{c}
H_1 \\ H_2 \\ H_3
\end{array}
\right)
=R
\left(
\begin{array}{c}
\eta_1 \\ \eta_2 \\ \eta_3
\end{array}
\right),
\end{equation}
where
\begin{equation} 
R=
\left(
\begin{array}{ccc}
R_{11} & R_{12} & R_{13} \\
R_{21}& R_{22} & R_{23} \\
R_{31} 
& R_{32} & R_{33}
\end{array}
\right),
\end{equation}
giving
\begin{equation}
\label{Eq:cal-M}
R{\cal M}^2R^{\rm T}={\rm diag}(M_1^2,M_2^2,M_3^2).
\end{equation}
The rotation matrix needed in order to diagonalize ${\cal M}^2$ will depend on the choice of basis. However, we must end up with the same physical fields, hence the same diagonalized mass matrix for any choice of basis. This implies that for two different choices of basis we must have
\bea
R
\left(
\begin{array}{c}
	\eta_1 \\ \eta_2 \\ \eta_3
\end{array}
\right)
=
\bar{R}
\left(
\begin{array}{c}
	\bar{\eta}_1 \\ \bar{\eta}_2 \\ \bar{\eta}_3
\end{array}
\right).
\eea
Since we know how the doublets $\Phi_i$ transform under a change of basis, we can use this to figure out how the $\eta_i$ transform under a change of basis. Combining this with the above, we arrive at the transformation rules for the rotation matrix $R$,
\bea
\bar{R}=RP,
\eea
where $P$ is an orthogonal 3x3 matrix with elements given by 
\begin{eqnarray}
P_{11}&=&\frac{\cos\theta (v_1 \cos\theta+v_2  \sin\theta\cos(\xi_{21}-\xi))}{\bar{v}_1},\\
P_{12}&=&-\frac{\sin\theta (v_2 \cos\theta \cos(\xi_{21}-\xi)-v_1 \sin\theta)}{\bar{v}_2},\\
P_{13}&=&\frac{v v_2 \sin 2\theta\sin(\xi_{21}-\xi)}{2 \bar{v}_1 \bar{v}_2},\\
P_{21}&=&\frac{\sin\theta (v_1 \cos\theta \cos(\xi_{21}-\xi)+v_2 \sin\theta)}{\bar{v}_1},\\
P_{22}&=&\frac{\cos\theta (v_2 \cos\theta-v_1 \sin\theta \cos(\xi_{21}-\xi) )}{\bar{v}_2},\\
P_{23}&=&-\frac{v v_1 \sin 2\theta\sin(\xi_{21}-\xi)}{2 \bar{v}_1 \bar{v}_2},\\
P_{31}&=&-\frac{v \sin 2\theta\sin(\xi_{21}-\xi)}{2 \bar{v}_1},\\
P_{32}&=&\frac{v \sin 2\theta\sin(\xi_{21}-\xi)}{2 \bar{v}_2},\\
P_{33}&=&\frac{2 v_1 v_2 \cos 2\theta+\left(v_2^2-v_1^2\right) \sin 2\theta \cos(\xi_{21}-\xi)}{2 \bar{v}_1 \bar{v}_2}.
\end{eqnarray}
By equating the diagonalized mass matrices for two different choices of basis,
\bea
R{\cal M}^2R^{\rm T}=\bar{R}\bar{{\cal M}}^2\bar{R}^{\rm T},
\eea
we find the transformation rules for the squared mass matrix under a basis transformation,
\bea
\bar{{\cal M}}^2&=&P^T{\cal M}^2P.
\eea
We make note of the fact that none of the elements of the rotation matrix or the squared mass matrix are invariant under a change of basis, so none of their elements are physical. By combining elements of the squared mass matrix like the trace, the sum of principal cofactors or the determinant we find basis invariant expressions, meaning that those combinations are physical. Indeed, we know that they can be expressed in terms of the three eigenvalues (masses) of the neutral sector. The characteristic equation of the squared mass matrix is
\bea
\lambda^3+b\lambda^2+c\lambda+d=0,
\eea 
with
\begin{eqnarray}
b&=&-({\cal M}_{11}^2+{\cal M}_{22}^2+{\cal M}_{33}^2),\\
c&=&{\cal M}_{11}^2{\cal M}_{22}^2+{\cal M}_{11}^2{\cal M}_{33}^2+{\cal M}_{22}^2{\cal M}_{33}^2-({\cal M}_{12}^2)^2-({\cal M}_{13}^2)^2-({\cal M}_{23}^2)^2,\\
d&=&{\cal M}_{11}^2({\cal M}_{23}^2)^2+{\cal M}_{22}^2({\cal M}_{13}^2)^2+{\cal M}_{33}^2({\cal M}_{12}^2)^2-{\cal M}_{11}^2{\cal M}_{22}^2{\cal M}_{33}^2-2{\cal M}_{12}^2{\cal M}_{13}^2{\cal M}_{23}^2.
\end{eqnarray}
Here, $a$, $b$ and $c$ are now basis invariant expressions. Introducing the (also basis invariant) combinations $p=c-b^2/3$ and $q=(2b^3-9bc+27d)/27$, we may write the three solutions of the characteristic equations as
\begin{eqnarray}
M_1^2&=&\frac{-b}{3}+2\sqrt{\frac{-p}{3}}\cos\left[\frac{1}{3}\arccos\left(\frac{3q}{2p}\sqrt{\frac{-3}{p}}\right)+\frac{2\pi}{3}\right],\\
M_2^2&=&\frac{-b}{3}+2\sqrt{\frac{-p}{3}}\cos\left[\frac{1}{3}\arccos\left(\frac{3q}{2p}\sqrt{\frac{-3}{p}}\right)-\frac{2\pi}{3}\right],\\
M_3^2&=&\frac{-b}{3}+2\sqrt{\frac{-p}{3}}\cos\left[\frac{1}{3}\arccos\left(\frac{3q}{2p}\sqrt{\frac{-3}{p}}\right)\right].
\end{eqnarray}
Since the three neutral masses can be written as functions of basis invariant quantities, this means that the neutral masses must also be basis invariant, again as we expect them to be. 
\section{Some important couplings and basis invariance}
\label{sec:Couplings}
\subsection{Scalar couplings}
We shall find the couplings between one neutral and two charged scalars as well as the coupling between four charged scalars useful. We can read the coefficients of these combinations of fields directly off of the potential. In the general basis they become
\begin{eqnarray}
q_{i}&\equiv&{\rm Coefficient}(V,H_iH^-H^+)\label{eq:qi}
\nonumber\\
&=&\frac{v_1v_2^2}{v^2}R_{i1}\lambda_1
+\frac{v_1^2v_2}{v^2}R_{i2}\lambda_2
+\frac{v_1^3R_{i1}+v_2^3R_{i2}}{v^2}\lambda_3
-\frac{v_1v_2(v_2R_{i1}+v_1R_{i2})}{v^2}\lambda_4\nonumber\\
&&
-\frac{v_1v_2\left[(v_2R_{i1}+v_1R_{i2})\cos2\xi_{21}-vR_{i3}\sin2\xi_{21}\right]}{v^2}
\Re\lambda_5\nonumber\\
&&
+\frac{v_1v_2\left[(v_2R_{i1}+v_1R_{i2})\sin2\xi_{21}+vR_{i3}\cos2\xi_{21}\right]}{v^2}\Im\lambda_5\nonumber\\
&&
+\frac{v_2\left(\left[(v_2^2-2v_1^2)R_{i1}+v_1v_2R_{i2}\right]\cos\xi_{21}-vv_2R_{i3}\sin\xi_{21}\right)}{v^2}\Re\lambda_6\nonumber\\
&&
-\frac{v_2\left(\left[(v_2^2-2v_1^2)R_{i1}+v_1v_2R_{i2}\right]\sin\xi_{21}+vv_2R_{i3}\cos\xi_{21}\right)}{v^2}\Im\lambda_6\nonumber\\
&&
+\frac{v_1\left(\left[(v_1^2-2v_2^2)R_{i2}+v_1v_2R_{i1}\right]\cos\xi_{21}-vv_1R_{i3}\sin\xi_{21}\right)}{v^2}\Re\lambda_7\nonumber\\
&&
-\frac{v_1\left(\left[(v_1^2-2v_2^2)R_{i2}+v_1v_2R_{i1}\right]\sin\xi_{21}+vv_1R_{i3}\cos\xi_{21}\right)}{v^2}\Im\lambda_7
,\\
q&\equiv&{\rm Coefficient}(V,H^-H^-H^+H^+)\label{eq:q}
\nonumber\\
&=&
\frac{v_2^4}{2v^4}\lambda_1
+\frac{v_1^4}{2v^4}\lambda_2
+\frac{v_1^2v_2^2}{v^4}(\lambda_3+\lambda_4+\cos2\xi_{21}\Re\lambda_5-\sin2\xi_{21}\Im\lambda_5)\nonumber\\
&&
-\frac{2v_1v_2^3}{v^4}(\cos\xi_{21}\Re\lambda_6-\sin\xi_{21}\Im\lambda_6)
-\frac{2v_1^3v_2}{v^4}(\cos\xi_{21}\Re\lambda_7-\sin\xi_{21}\Im\lambda_7).
\end{eqnarray}
Again, upon investigating how these expressions transform under a change of basis, we conclude that both the $q_i$ and $q$ are basis invariant expressions. Hence, they are physical and can be measured in experiments. Again, this is as it should be, as the couplings of the theory appear in the Feynman amplitudes and are directly related to observables. 
\subsection{Gauge couplings}
The kinetic term of the Lagrangian in the 2HDM may be written
\begin{equation} \label{Eq:gauge-IS}
\lcal_k=(D_\mu \Phi_1)^\dagger(D^\mu \Phi_1) + (D_\mu \Phi_2)^\dagger(D^\mu \Phi_2)
,
\end{equation}
with 
\bea
D^\mu&=&\partial^\mu+\frac{ig}{2}\sigma_iW_i^\mu+i\frac{g^\prime}{2}B^\mu,\\
W_1^\mu&=&\frac{1}{\sqrt{2}}(W^{+\mu}+W^{-\mu}),\\
W_2^\mu&=&\frac{i}{\sqrt{2}}(W^{+\mu}-W^{-\mu}),\\
W_3^\mu&=&\cos\thetaW Z^\mu+\sin\thetaW A^\mu,\\
B^\mu&=&-\sin\thetaW Z^\mu+\cos\thetaW A^\mu.
\eea
The couplings between scalars and gauge bosons can now be read directly off from the kinetic term, i.e.
\bea
{\rm Coefficient}\left(\lcal_k,Z^\mu \left[H_j \overleftrightarrow{\partial_\mu} H_i\right]\right)&=&\frac{g}{2v\cos\thetaW}\epsilon_{ijk}e_k,\\
{\rm Coefficient}\left(\lcal_k,H_i Z^\mu Z^\nu\right)&=&\frac{g^2}{4\cos^2\thetaW}e_i\,g_{\mu\nu},\\
{\rm Coefficient}\left(\lcal_k,H_i W^{+\mu} W^{-\nu}\right)&=&\frac{g^2}{2}e_i\,g_{\mu\nu}.
\eea
The factors $e_i$ appearing in these couplings are given by
\begin{eqnarray}
e_i\equiv v_1R_{i1}+v_2R_{i2},
\end{eqnarray}
and satisfy $e_1^2+e_2^2+e_3^3=v^2=(246\, {\rm GeV})^2$. Upon performing a change of basis, we easily find that $\bar{v}_1\bar{R}_{i1}+\bar{v}_2\bar{R}_{i2}=v_1R_{i1}+v_2R_{i2}$, implying that the $e_i$ are basis invariant quantities. Hence, these gauge couplings are also physical and can be measured in experiments.
\section{Systematic construction of invariants}
\label{sec:construction}
Invariants can also be constructed in a systematical way by using tensors and their transformation properties under a change of basis. We have already encountered the $Y$- and $Z$-tensors. Let us also introduce the $V$-tensor defined by
\begin{eqnarray}
V_{ab}&=&
\frac{v_av_b^*}{v^2}
=
\frac{1}{v^2}\left(
\begin{array}{ccc}
v_1^2 & v_1 v_2e^{-i \xi_{21}}    \\
v_1 v_2e^{i \xi_{21}} & v_2^2  \\

\end{array}
\right).
\end{eqnarray}
Under a change of basis, the transformation rules of the tensors can be found in \cite{Gunion:2005ja} and are given by\footnote{The barred indices can be used to see which indices transform with $U$ and which transforms with $U^\dagger$.}
\begin{eqnarray}
\bar{Y}_{a\bar{b}}&=&U_{a\bar{c}}Y_{c\bar{d}}U_{d\bar{b}}^\dagger,\\
\bar{V}_{a\bar{b}}&=&U_{a\bar{c}}V_{c\bar{d}}U_{d\bar{b}}^\dagger,\\
\bar{Z}_{a\bar{b}c\bar{d}}&=&U_{a\bar{e}}U_{c\bar{g}}Z_{e\bar{f}g\bar{h}}U_{f\bar{b}}^\dagger U_{h\bar{d}}^\dagger.
\end{eqnarray}
In order to construct a basis invariant from these three tensors we must write down an expression that does not depend on the unitary matrix $U$. We put together an arbitrary number of $Y$-, $V$- and $Z$-tensors, pairing together barred and unbarred indices that are not already summed over. Next, we sum over the indices we pair together, exploiting the unitarity of $U$ to get an expression that is independent of $U$, hence basis invariant. The simplest examples are
\begin{eqnarray}
V_{a\bar{a}}&=&1,\\
Y_{a\bar{a}}&=&-\frac{1}{2}(m_{11}^2+m_{22}^2),\\
Z_{a\bar{a}b\bar{b}}&=&\lambda_1+\lambda_2+2\lambda_3,\\
Z_{a\bar{b}b\bar{a}}&=&\lambda_1+\lambda_2+2\lambda_4.
\end{eqnarray}
We recognize those basis invariant expressions that we already encountered in eqs. (\ref{ind1}) - (\ref{ind3}).
\section{From parameters to physical quantities}
\label{sec:translation}
We have  seen that the masses of the scalars as well as some couplings involving scalars are all invariant under a change of basis. As for invariants constructed by contracting barred and unbarred indices from the tensors, our desire is to express these in terms of masses and couplings, and thereby relating them to physical quantities. It has been pointed out by several groups \cite{Davidson:2005cw,Olaussen:2010aq} that the 2HDM has only 11 independent physical parameters. This is consistent with our findings, that all basis invariants can be expressed in terms of the following set of masses and couplings already encountered,
\bea
{\cal P}\equiv\{M_{H^\pm}^2,M_1^2,M_2^2,M_3^2,e_1,e_2,e_3,q_1,q_2,q_3,q\}.
\eea
We have explicitly shown that all masses and couplings are basis invariant. The challenge ahead of us is to find a way to translate a basis invariant expression into a combination of masses and couplings contained in the physical parameter set $\pcal$. The method developed uses the freedom to choose a particular basis, namely the Higgs-basis in which only the first doublet has a non-zero VEV, which is real.
\begin{equation}
\langle \Phi_1 \rangle_{\text{HB}}=\frac{1}{\sqrt{2}}\left(
\begin{array}{c}0\\
v

\end{array}\right),\quad
\langle \Phi_2 \rangle_{\text{HB}} =\left(
\begin{array}{c}0\\
0
\end{array}\right)\nonumber
\end{equation}
Next, we find the relation between parameters and physical quantities we seek in this particular basis. Finally, we use the fact that the quantities we translated into a combination of physical masses/couplings is basis invariant, hence the relations we found in the Higgs-basis must be valid in any basis.
\subsection{The translation}
We start by working out the stationary-point equations in the Higgs-basis. They become
\begin{eqnarray}
Y_{11}&=&-\frac{v^2}{2}Z_{1111},\\
\Re Y_{12}&=&-\frac{v^2}{2}\Re Z_{1112},\\
\Im Y_{12}&=&-\frac{v^2}{2}\Im Z_{1112}.
\end{eqnarray}
Next, we find the masses of the charged scalars,
\begin{eqnarray}
M_{H^\pm}^2=Y_{22}+\frac{v^2}{2} Z_{1122}.\label{chargedmass}
\end{eqnarray} 
The neutral-sector squared mass matrix is given by
\begin{eqnarray} 
&&{\cal M}^2
=R^T{\rm diag}(M_1^2,M_2^2,M_3^2)R\nonumber\\
&&=v^2
\begin{pmatrix}
Z_{1111} & \Re Z_{1112} & -\Im Z_{1112} \\
\Re Z_{1112}& \frac{1}{2}(Z_{1122}+Z_{1221}+\Re Z_{1212})+\frac{Y_{22}}{v^2} & -\frac{1}{2}\Im Z_{1212} \\
-\Im Z_{1112} 
& -\frac{1}{2}\Im Z_{1212} & \frac{1}{2}(Z_{1122}+Z_{1221}-\Re Z_{1212})+\frac{Y_{22}}{v^2}
\end{pmatrix}.\label{neutralmassmatrix}
\end{eqnarray}
We treat (\ref{chargedmass}) and (\ref{neutralmassmatrix}) as seven equations and solve them to get
\begin{eqnarray} 
Y_{22}&=&M_{H^\pm}^2-\frac{v^2}{2}Z_{1122},\\
Z_{1111}&=&\frac{R_{11}^2 M_1^2+R_{21}^2 M_2^2+R_{31}^2 M_3^2}{v^2},\\
Z_{1221}&=&\frac{-2 M_{H^\pm}^2+(R_{12}^2+R_{13}^2) M_1^2+(R_{22}^2+R_{23}^2) M_2^2+(R_{32}^2+R_{33}^2) M_3^2}{v^2},\\
\Re Z_{1112}&=&\frac{R_{11}R_{12} M_1^2+R_{21}R_{22} M_2^2+R_{31}R_{32} M_3^2}{v^2},\\
\Im Z_{1112}&=&-\frac{R_{11}R_{13} M_1^2+R_{21}R_{23} M_2^2+R_{31}R_{33} M_3^2}{v^2},\\
\Re Z_{1212}&=&\frac{(R_{12}^2-R_{13}^2) M_1^2+(R_{22}^2-R_{23}^2) M_2^2+(R_{32}^2-R_{33}^2) M_3^2}{v^2},\\
\Im Z_{1212}&=&-2\frac{R_{12}R_{13} M_1^2+R_{22}R_{23} M_2^2+R_{32}R_{33} M_3^2}{v^2}.
\end{eqnarray}
Next, we work out the scalar couplings to find
\begin{eqnarray}
q_i&=&v(R_{i1}Z_{1122}+R_{i2}\Re Z_{1222}-R_{i3}\Im Z_{1222}),\\
q&=&\frac{1}{2}Z_{2222}.\label{scalarcouplings}
\end{eqnarray}
We treat these as four equations and solve to get
\begin{eqnarray}
Z_{1122}&=&\frac{R_{11}q_1+R_{21}q_2+R_{31}q_3}{v},\\
\Re Z_{1222}&=&\frac{R_{12}q_1+R_{22}q_2+R_{32}q_3}{v},\\
\Im Z_{1222}&=&-\frac{R_{13}q_1+R_{23}q_2+R_{33}q_3}{v},\\
Z_{2222}&=&2q.
\end{eqnarray} 
We are now able to replace all the parameters of the potential with masses, scalar couplings and elements from the rotation matrix $R$. It turns out that all the combinations of rotation-matrix elements appearing in basis invariants during this translation is expressible in terms of the three elements $R_{i1}$ of the first column using the fact that $R$ is orthogonal. Finally, we turn to the factors appearing in the gauge couplings. In the Higgs-basis these are simply $e_i=vR_{i1}$, so indeed we manage to express all invariants in terms of the physical parameters of $\pcal$.

Applying the translation technique oulined above, we find
\begin{eqnarray}
Y_{a\bar{a}}&=&M_{H^{\pm}}^2-\frac{1}{2v^2}(e_1^2M_1^2+e_2^2M_2^2+e_3^2M_3^2)-\frac{1}{2}(e_1q_1+e_2q_2+e_3q_3),\nonumber\\
Z_{a\bar{a}b\bar{b}}&=&2q+\frac{1}{v^4}(e_1^2M_1^2+e_2^2M_2^2+e_3^2M_3^2)+\frac{2}{v^2}(e_1q_1+e_2q_2+e_3q_3),\nonumber\\
Z_{a\bar{b}b\bar{a}}&=&2q-\frac{4}{v^2}M_{H^{\pm}}^2-\frac{1}{v^4}(e_1^2M_1^2+e_2^2M_2^2+e_3^2M_3^2)+\frac{2}{v^2}(M_1^2+M_2^2+M_3^2),\nonumber
\end{eqnarray}
all now translated into masses and couplings.
\section{Applying the technique to CP odd invariants}
\label{sec:invariants}
CP violation in the bosonic sector of the 2HDM has been extensively studied by many groups with different approaches \cite{Lavoura:1994fv,Botella:1994cs,Gunion:2005ja,Branco:2005em,Ivanov:2005hg,Nishi:2006tg,Ivanov:2006yq,Maniatis:2007vn,Branco:1999fs,El_Kaffas:2006nt,Grzadkowski:2013rza}.
The three invariants we translated at the end of the previous section were all CP even since they contain no imaginary part. In \cite{Gunion:2005ja}, the conditions for CP conservation were formulated in terms of three CP odd invariants expressed using the tensors $Y$, $V$ and $Z$. If at least one of these is non-zero, then the 2HDM violates CP, and if all three invariants vanish it is CP conserving. The three invariants are
\begin{subequations}
	\begin{align}
	\Im J_1&=-\frac{2}{v^2}\Im\bigl[V_{d\bar{a}} Y_{a\bar{b}} Z_{b\bar{c}c\bar{d}}\bigr], \\
	\Im J_2&=\frac{4}{v^4}\Im\bigl[V_{a\bar{b}}V_{d\bar{c}} Y_{b\bar{e}} Y_{c\bar{f}} Z_{e\bar{a}f\bar{d}}\bigr],\\
	\Im J_3&=\Im\bigl[V_{a\bar{b}}V_{d\bar{c}} Z_{b\bar{g}g\bar{e}} Z_{c\bar{h}h\bar{f}}Z_{e\bar{a}f\bar{d}}\bigr].
	\label{eq:im_J3}
	\end{align}
\end{subequations}
We translate these using the technique described to get
\begin{eqnarray}
\Im J_1&=&\frac{1}{v^5}\sum_{i,j,k}\epsilon_{ijk}M_i^2e_ie_kq_j,\\
\Im J_2&=&2\frac{e_1 e_2 e_3}{v^9}(M_1^2-M_2^2)(M_2^2-M_3^2)(M_3^2-M_1^2),\\
\Im J_3&=& \frac{2}{v^4}\left[ (e_1^2M_1^2+e_2^2M_2^2+e_3^2M_3^2)
+v^2(e_1q_1  + e_2q_2 +  e_3q_3) +2v^2M_{H^\pm}^2\right]\Im J_1\nonumber\\
&& +\Im J_2+\frac{4}{v^7}\sum_{i,j,k}\epsilon_{ijk}e_i M_i^4  e_j q_k+2\Im J_{30},
\end{eqnarray}
where
\bea
\Im J_{30}&=&\frac{1}{v^5}\sum_{i,j,k}\epsilon_{ijk} q_i M_i^2  e_j q_k.
\eea
These results were obtained and presented in \cite{Grzadkowski:2014ada}, where a more complicated technique than the one presented here was used. 
If $\Im J_1=\Im J_2=0$, we find that $\Im J_3=\Im J_{30}$, so a simpler way to express the conditions for CP conservation is to say that the model conserves CP iff $\Im J_1=\Im J_2=\Im J_{30}=0$. Othervise it violates CP. Putting $\Im J_1=\Im J_2=\Im J_{30}=0$, we immediately find the following six cases of CP conservation:
\\
Case 1: $M_1=M_2=M_3$. Full mass deceneracy.\\
Case 2: $M_1=M_2$ and $e_1q_2 = e_2q_1$.\\
Case 3: $M_2=M_3$ and $e_2q_3 = e_3q_2$.\\
Case 4: $e_1=0$ and $q_1=0$.\\
Case 5: $e_2=0$ and $q_2=0$.\\
Case 6: $e_3=0$ and $q_3=0$.\\
\\
If none of these are satisfied, the model violates CP.

The technique can also be applied on the four invariants needed to determine if CP is violated explicitly or spontaneously presented in eqs. (23) - (26) of \cite{Gunion:2005ja}, arriving at much simpler conditions given in terms of masses and couplings, presented in eqs. (III.5) - (III.7) of \cite{Grzadkowski:2016szj}.
\section{Summary}
\label{sec:summary}
We have discussed the importance of basis invariants in the 2HDM and shown how the masses and some couplings of the theory are invariant under a change of basis. We have also discussed how to construct basis invariants in a systematic way by use of the $Y$-, $V$- and $Z$-tensors. We have presented a simple, yet powerful method to translate invariant expressions into combinations of 11 masses and couplings and demonstrated its powerfulness and simplicity by applying the method to some CP-even invariants as well as the three CP-odd invariants needed to determine the CP properties of the theory. The technique can be applied to any invariant expression of the 2HDM. This technique could be extended also to the three-Higgs-doublet model in order to study the CP properties of the 3HDM.


\begin{thebibliography}{99}
\bibitem{Branco:2011iw}
G.~C.~Branco, P.~M.~Ferreira, L.~Lavoura, M.~N.~Rebelo, M.~Sher and J.~P.~Silva,
{\it Theory and phenomenology of two-Higgs-doublet models,}
Phys.\ Rept.\  {\bf 516} (2012) 1
[arXiv:1106.0034 [hep-ph]].


\bibitem{Davidson:2005cw}
S.~Davidson and H.~E.~Haber,
{\it Basis-independent methods for the two-Higgs-doublet model,}
Phys.\ Rev.\ D {\bf 72} (2005) 035004
[Erratum-ibid.\ D {\bf 72} (2005) 099902]
[hep-ph/0504050].

\bibitem{Haber:2006ue} 
H.~E.~Haber and D.~O'Neil,
Phys.\ Rev.\ D {\bf 74}, 015018 (2006)
[hep-ph/0602242].

\bibitem{Haber:2010bw}
H.~E.~Haber and D.~O'Neil,
Phys.\ Rev.\ D {\bf 83} (2011) 055017
doi:10.1103/PhysRevD.83.055017
[arXiv:1011.6188 [hep-ph]].

\bibitem{Gunion:2005ja}
J.~F.~Gunion and H.~E.~Haber,
{\it Conditions for CP-violation in the general two-Higgs-doublet model,}
Phys.\ Rev.\ D {\bf 72} (2005) 095002
[hep-ph/0506227].

\bibitem{Grzadkowski:2014ada}
B.~Grzadkowski, O.~M.~Ogreid and P.~Osland,
JHEP {\bf 1411} (2014) 084
doi:10.1007/JHEP11(2014)084
[arXiv:1409.7265 [hep-ph]].

\bibitem{Olaussen:2010aq}
K.~Olaussen, P.~Osland and M.~A.~Solberg,
JHEP {\bf 1107} (2011) 020
doi:10.1007/JHEP07(2011)020
[arXiv:1007.1424 [hep-ph]].


\bibitem{Lavoura:1994fv}
L.~Lavoura and J.~P.~Silva,
{\it Fundamental CP violating quantities in a SU(2) $\times$ U(1) model with many Higgs doublets,}
Phys.\ Rev.\ D {\bf 50} (1994) 4619
[hep-ph/9404276].

\bibitem{Botella:1994cs}
F.~J.~Botella and J.~P.~Silva,
{\it Jarlskog-like invariants for theories with scalars and fermions,}
Phys.\ Rev.\ D {\bf 51} (1995) 3870
[hep-ph/9411288].

\bibitem{Branco:2005em} 
G.~C.~Branco, M.~N.~Rebelo and J.~I.~Silva-Marcos,
{\it CP-odd invariants in models with several Higgs doublets,}
Phys.\ Lett.\ B {\bf 614}, 187 (2005)
[hep-ph/0502118].



\bibitem{Ivanov:2005hg}
I.~P.~Ivanov,
{\it Two-Higgs-doublet model from the group-theoretic perspective,}
Phys.\ Lett.\ B {\bf 632} (2006) 360
[hep-ph/0507132].

\bibitem{Nishi:2006tg}
C.~C.~Nishi,
{\it CP violation conditions in N-Higgs-doublet potentials,}
Phys.\ Rev.\ D {\bf 74} (2006) 036003
[Erratum-ibid.\ D {\bf 76} (2007) 119901]
[hep-ph/0605153].

\bibitem{Ivanov:2006yq}
I.~P.~Ivanov,
{\it Minkowski space structure of the Higgs potential in 2HDM,}
Phys.\ Rev.\ D {\bf 75} (2007) 035001
[Erratum-ibid.\ D {\bf 76} (2007) 039902]
[hep-ph/0609018].

\bibitem{Maniatis:2007vn}
M.~Maniatis, A.~von Manteuffel and O.~Nachtmann,
{\it CP violation in the general two-Higgs-doublet model: A Geometric view,}
Eur.\ Phys.\ J.\ C {\bf 57} (2008) 719
[arXiv:0707.3344 [hep-ph]].

\bibitem{Branco:1999fs} 
G.~C.~Branco, L.~Lavoura and J.~P.~Silva,
{\it CP Violation,}
Int.\ Ser.\ Monogr.\ Phys.\  {\bf 103}, 1 (1999).

\bibitem{El_Kaffas:2006nt}
A.~W.~El Kaffas, W.~Khater, O.~M.~Ogreid and P.~Osland,
{\it Consistency of the two Higgs doublet model and CP violation in top production at the LHC,}
Nucl.\ Phys.\ B {\bf 775} (2007) 45
[hep-ph/0605142].


\bibitem{Grzadkowski:2013rza}
B.~Grzadkowski, O.~M.~Ogreid and P.~Osland,
{\it Diagnosing CP properties of the 2HDM,}
JHEP {\bf 1401} (2014) 105
[arXiv:1309.6229 [hep-ph], arXiv:1309.6229].

\bibitem{Grzadkowski:2016szj}
B.~Grzadkowski, O.~M.~Ogreid and P.~Osland,
Phys.\ Rev.\ D {\bf 94} (2016) no.11,  115002
doi:10.1103/PhysRevD.94.115002
[arXiv:1609.04764 [hep-ph]].


\end{thebibliography}
\end{document}